\documentclass[12pt]{article}

\usepackage{axodraw}
\usepackage{epsf}
\usepackage{rotating}
\usepackage{a4wide}
\usepackage{graphics}

\catcode`@=11
\def\citer{\@ifnextchar
[{\@tempswatrue\@citexr}{\@tempswafalse\@citexr[]}}

\def\@citexr[#1]#2{\if@filesw\immediate\write\@auxout{\string\citation{#2}}\fi
  \def\@citea{}\@cite{\@for\@citeb:=#2\do
    {\@citea\def\@citea{--\penalty\@m}\@ifundefined
       {b@\@citeb}{{\bf ?}\@warning
       {Citation `\@citeb' on page \thepage \space undefined}}%
\hbox{\csname b@\@citeb\endcsname}}}{#1}}
\catcode`@=12

\newcommand{\beq}{\begin{equation}}
\newcommand{\eeq}{\end{equation}}
\newcommand{\bea}{\begin{eqnarray}}
\newcommand{\eea}{\end{eqnarray}}
\newcommand{\bed}{\begin{displaymath}}
\newcommand{\eed}{\end{displaymath}}

\newcommand{\non}{\nonumber}
\newcommand{\sia}{\sin\alpha}
\newcommand{\coa}{\cos\alpha}
\newcommand{\sib}{\sin\beta}
\newcommand{\cob}{\cos\beta}
\newcommand{\tga}{{\rm tg}\alpha}
\newcommand{\tgb}{{\rm tg}\beta}
\newcommand{\stgb}{{\rm tg}^2\beta}
\newcommand{\sq}{\tilde{b}}
\newcommand{\sgl}{\tilde{g}}
\newcommand{\MS}{\overline{\rm MS}}

\def\sla#1{\ifmmode% 
\setbox0=\hbox{$#1$}% 
\setbox1=\hbox to\wd0{\hss$/$\hss}\else% 
\setbox0=\hbox{#1}% 
\setbox1=\hbox to\wd0{\hss/\hss}\fi% 
#1\hskip-\wd0\box1 } 

\newcommand{\lsim}{\raisebox{-0.13cm}{~\shortstack{$<$ \\[-0.07cm] $\sim$}}~}

\newcommand{\mb}{\ensuremath{m_b}}
\newcommand{\Dmb}{\ensuremath{\Delta \mb}}

\newcommand{\mg}{\ensuremath{m_{\tilde g}}}

\newcommand{\figone}{\begin{figure}[htb]
\centerline{\includegraphics{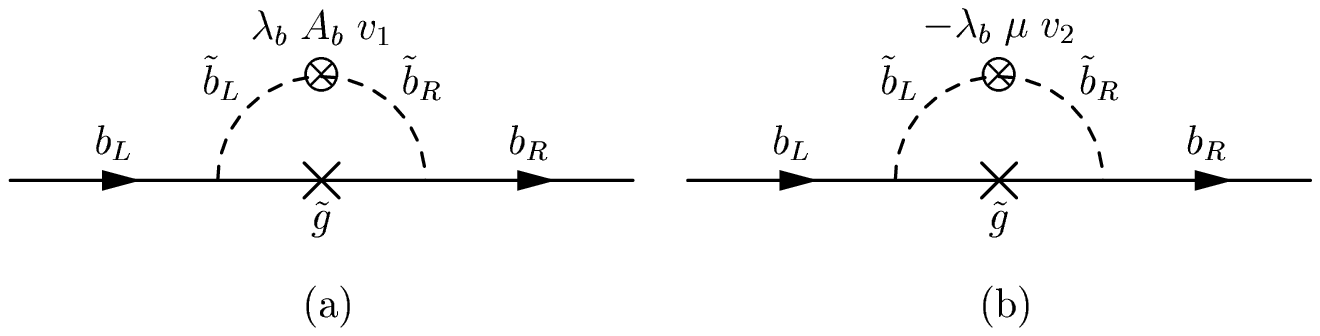}}
\caption{One-loop contribution to the quantities \textbf{(a)} $\Delta_1$ and
\textbf{(b)} $\Delta_2$.}
\label{fg:leading}
\end{figure}}

\def\draftdate{\relax}
\def\mda{\relax}
\def\mua{\relax}
\def\mla{\relax}
\def\draft{
\def\thtystars{******************************}
\def\sixtystars{\thtystars\thtystars}
\typeout{}
\typeout{\sixtystars**}
\typeout{* Draft mode!
         For final version remove \protect\draft\space in source file *}
\typeout{\sixtystars**}
\typeout{}
\def\draftdate{\today}
\def\mua{\marginpar[\boldmath\hfil$\uparrow$]%
                   {\boldmath$\uparrow$\hfil}%
                    \typeout{marginpar: $\uparrow$}\ignorespaces}
\def\mda{\marginpar[\boldmath\hfil$\downarrow$]%
                   {\boldmath$\downarrow$\hfil}%
                    \typeout{marginpar: $\downarrow$}\ignorespaces}
\def\mla{\marginpar[\boldmath\hfil$\rightarrow$]%
                   {\boldmath$\leftarrow $\hfil}%
                    \typeout{marginpar: $\leftrightarrow$}\ignorespaces}
\def\Mua{\marginpar[\boldmath\hfil$\Uparrow$]%
                   {\boldmath$\Uparrow$\hfil}%
                    \typeout{marginpar: $\uparrow$}\ignorespaces}
\def\Mda{\marginpar[\boldmath\hfil$\Downarrow$]%
                   {\boldmath$\Downarrow$\hfil}%
                    \typeout{marginpar: $\downarrow$}\ignorespaces}
\def\Mla{\marginpar[\boldmath\hfil$\Rightarrow$]%
                   {\boldmath$\Leftarrow $\hfil}%
                    \typeout{marginpar: $\leftrightarrow$}\ignorespaces}
\overfullrule 5pt
\oddsidemargin -15mm
\marginparwidth 29mm
}

\def\stars{\strut\leaders\hbox{*}\hfill\strut}
\def\starline{\hfil\strut\hfil\hbox to \textwidth {\stars}\hfil}

\catcode`\@=11
\ifx\@Hxfloat\@Hundef\else\expandafter\endinput\fi
\let\@Hxfloat\@xfloat
\def\@xfloat#1[{\@ifnextchar{H}{\@HHfloat{#1}[}{\@Hxfloat{#1}[}}
\def\@HHfloat#1[H]{%
\expandafter\let\csname end#1\endcsname\end@Hfloat
\vskip\intextsep\vbox\bgroup\def\@captype{#1}\parindent\z@
\ignorespaces}
\def\end@Hfloat{\egroup\vskip \intextsep}
\relax

\begin{document}
 
\renewcommand{\thefootnote}{\fnsymbol{footnote} }

\vskip-1.0cm

\begin{flushright}
PSI--PR--03--06 \\
hep-ph/0305101
\end{flushright}

\begin{center}
{\large\sc MSSM Higgs Decays to Bottom Quark Pairs revisited}%
\footnote{This work has been supported in part by the Swiss Bundesamt
f\"ur Bildung und Wissenschaft and by the European Union under contract
HPRN-CT-2000-00149.}
\end{center}

\begin{center}
Jaume Guasch$^1$, Petra H\"afliger$^{1,2}$ and Michael Spira$^1$
\end{center}

%\vspace*{0.4cm}

\begin{center}
{\it \small
$^1$ Paul Scherrer Institut, CH-5232 Villigen PSI, Switzerland \\
$^2$ Institute for Particle Physics, ETH Z\"urich, CH-8093 Z\"urich,
Switzerland}
\end{center}
%\vspace*{0.0cm}

\begin{abstract}
We present an update of neutral Higgs boson decays into bottom quark
pairs in the minimal supersymmetric extension of the Standard Model. In
particular the resummation of potentially large higher-order corrections
due to the soft SUSY breaking parameters $A_b$ and $\mu$ is extended.
The remaining theoretical uncertainties due to unknown higher-order
SUSY-QCD corrections are analyzed quantitatively.
\end{abstract}

\def\thefootnote{\arabic{footnote}}
\setcounter{footnote}{0}

\section{Introduction}
%        ============
The Higgs mechanism is a cornerstone of the Standard Model (SM) and its
supersymmetric extensions. The search for Higgs bosons is one of the
most important endeavors at future high-energy experiments.  Since the
minimal supersymmetric extension of the Standard Model (MSSM) requires
the introduction of two Higgs doublets in order to preserve
supersymmetry, there are five elementary Higgs particles, two CP-even
($h,H$), one CP-odd ($A$) and two charged ones ($H^\pm$). At lowest
order all couplings and masses of the MSSM Higgs sector are fixed by two
independent input parameters, which are generally chosen as
$\tgb=v_2/v_1$, the ratio of the two vacuum expectation values
$v_{1,2}$, and the pseudoscalar Higgs-boson mass $M_A$. At LO the light
scalar Higgs mass $M_h$ has to be smaller than the $Z$-boson mass $M_Z$.
Including the one-loop and dominant two-loop corrections the upper bound
is increased to $M_h\lsim 135$ GeV \cite{mssmrad}.  The couplings of the
various neutral Higgs bosons to fermions and gauge bosons depend on the
angles $\alpha$ and $\beta$. Normalized to the SM Higgs couplings, they
are listed in Table~\ref{tb:hcoup}.
\begin{table}[hbt]
\renewcommand{\arraystretch}{1.5}
\begin{center}
\begin{tabular}{|lc||ccc|} \hline
\multicolumn{2}{|c||}{$\Phi$} & $g^\Phi_u$ & $g^\Phi_d$ &  $g^\Phi_V$ \\
\hline \hline
SM~ & $H$ & 1 & 1 & 1 \\ \hline
MSSM~ & $h$ & $\cos\alpha/\sin\beta$ & $-\sin\alpha/\cos\beta$ &
$\sin(\beta-\alpha)$ \\
& $H$ & $\sin\alpha/\sin\beta$ & $\cos\alpha/\cos\beta$ &
$\cos(\beta-\alpha)$ \\
& $A$ & $ 1/\tgb$ & $\tgb$ & 0 \\ \hline
\end{tabular}
\renewcommand{\arraystretch}{1.2}
\caption[]{\label{tb:hcoup}
\it Higgs couplings in the MSSM to fermions and gauge bosons [$V=W,Z$]
relative to SM couplings.}
\end{center}
\end{table}

The pseudoscalar particle $A$ does not couple to gauge bosons at tree
level, and its couplings to down (up)-type fermions are (inversely)
proportional to $\tgb$.  The negative direct searches for the
Higgsstrahlung processes $e^+e^-\to Zh,ZH$ and the associated production
$e^+e^-\to Ah,AH$ yield lower bounds of $M_{h,H} > 91.0$ GeV and $M_A >
91.9$ GeV. The range $0.5 < \tgb < 2.4$ in the MSSM is excluded by the
Higgs searches for a SUSY scale $M_{SUSY}=1$ TeV at the LEP2 experiments
\cite{lep2}\footnote{The excluded range of $\tgb$ values
depends significantly on the value of the top-quark
mass~\cite{Heinemeyer:1999zf}.}.

The scalar superpartners $\tilde f_{L,R}$ of the left- and right-handed
fermion
components mix with each other. The mass eigenstates $\tilde f_{1,2}$ of
the
sfermions $\tilde f$ are related to the current eigenstates $\tilde
f_{L,R}$
by mixing angles $\theta_f$,
\begin{eqnarray}
\tilde f_1 & = & \tilde f_L \cos\theta_f + \tilde f_R \sin \theta_f
\nonumber \\
\tilde f_2 & = & -\tilde f_L\sin\theta_f + \tilde f_R \cos \theta_f \, ,
\label{eq:sfmix}
\end{eqnarray}
which are proportional to the masses of the ordinary fermions. Thus
mixing
effects are only important for the third-generation sfermions $\tilde t,
\tilde
b, \tilde \tau$, the mass matrix of which is given by
\cite{mssmbase}\footnote{For simplicity, the $D$-terms have been absorbed
in the sfermion mass parameters $M_{\tilde f_{L/R}}^2$.}
\begin{equation}
{\cal M}_{\tilde f} = \left[ \begin{array}{cc}
M_{\tilde f_L}^2 + m_f^2 & m_f (A_f-\mu r_f) \\
m_f (A_f-\mu r_f) & M_{\tilde f_R}^2 + m_f^2
\end{array} \right] \, ,
\label{eq:sqmassmat}
\end{equation}
with the parameters $r_b = r_\tau = 1/r_t = \tgb$. The parameters $A_f$
denote the trilinear scalar coupling of the soft supersymmetry
breaking part of the Lagrangian. Consequently the mixing angles acquire
the form
\begin{equation}
\sin 2\theta_f = \frac{2m_f (A_f-\mu r_f)}{M_{\tilde f_1}^2 - M_{\tilde
f_2}^2}
~~~,~~~
\cos 2\theta_f = \frac{M_{\tilde f_L}^2 - M_{\tilde f_R}^2}{M_{\tilde
f_1}^2
- M_{\tilde f_2}^2}
\end{equation}
and the masses of the squark mass eigenstates are given by
\begin{equation}
M_{\tilde f_{1,2}}^2 = m_f^2 + \frac{1}{2}\left[ M_{\tilde f_L}^2 +
M_{\tilde f_R}^2 \mp \sqrt{(M_{\tilde f_L}^2 - M_{\tilde f_R}^2)^2 +
4m_f^2 (A_f - \mu r_f)^2} \right] \, .
\end{equation}
The neutral Higgs couplings to sfermions read as \cite{DSUSY}
\begin{eqnarray}
g_{\tilde f_L \tilde f_L}^\Phi & = & m_f^2 g_1^\Phi + M_Z^2 (I_{3f}
- e_f\sin^2\theta_W) g_2^\Phi \nonumber \\
g_{\tilde f_R \tilde f_R}^\Phi & = & m_f^2 g_1^\Phi + M_Z^2
e_f\sin^2\theta_W
g_2^\Phi \nonumber \\
g_{\tilde f_L \tilde f_R}^\Phi & = & -\frac{m_f}{2} (\mu g_3^\Phi
- A_f g_4^\Phi) \, ,
\label{eq:hsfcouprl}
\end{eqnarray}
with the couplings $g_i^\Phi$ listed in Table \ref{tb:hsfcoup}.
\begin{table}[hbt]
\renewcommand{\arraystretch}{1.5}
\begin{center}
\begin{tabular}{|l|c||c|c|c|c|} \hline
$\tilde f$ & $\Phi$ & $g^\Phi_1$ & $g^\Phi_2$ & $g^\Phi_3$ & $g^\Phi_4$
\\
\hline \hline
& $h$ & $\cos\alpha/\sin\beta$ & $-\sin(\alpha+\beta)$ &
$-\sin\alpha/\sin\beta$ & $\cos\alpha/\sin\beta$ \\
$\tilde u$ & $H$ & $\sin\alpha/\sin\beta$ & $\cos(\alpha+\beta)$ &
$\cos\alpha/\sin\beta$ & $\sin\alpha/\sin\beta$ \\
& $A$ & 0 & 0 & $-1$ & $1/\tgb$ \\ \hline
& $h$ & $-\sin\alpha/\cos\beta$ & $-\sin(\alpha+\beta)$ &
$\cos\alpha/\cos\beta$ & $-\sin\alpha/\cos\beta$ \\
$\tilde d$ & $H$ & $\cos\alpha/\cos\beta$ & $\cos(\alpha+\beta)$ &
$\sin\alpha/\cos\beta$ & $\cos\alpha/\cos\beta$ \\
& $A$ & 0 & 0 & $-1$ & $\tgb$ \\ \hline
\end{tabular}
\renewcommand{\arraystretch}{1.2}
\caption[]{\label{tb:hsfcoup}
\it Coefficients of the neutral MSSM Higgs couplings to sfermion pairs.}
\end{center}
\end{table}

In this paper we investigate the theoretical status of SUSY--QCD
corrections to neutral Higgs decays into bottom quark pairs. In
particular we concentrate on the theoretical uncertainties of the
partial width in regions, where the SUSY--QCD corrections are large,
i.e.\,for large values of $\tgb$ and sizeable magnitudes of the Higgsino
mass parameter $\mu$ \cite{dmbbase}. These regions are particularly
interesting, since the contributions generated by gluino exchange are
enhanced by $\tgb$. They play an important role in the phenomenology of
SUSY-Higgs bosons at high-energy colliders, since they shift the
Higgs-boson discovery and exclusion regions significantly \cite{hphen}.
The corrections can also provide at distinction between supersymmetric
and non-supersymmetric Higgs bosons.  The dominant contributions have
been resummed before \cite{CGNW}. However, the trilinear coupling $A_b$
may be large, too. We extend the resummation by including the dominant
$A_b$ terms. 

Although we investigate only the SUSY-QCD corrections, it
should be noted that the electroweak corrections can be important, too,
and yield an additional contribution to the uncertainties. The full
one-loop electroweak corrections were computed in Ref.~\cite{dabel}, and
later refined in~\cite{Heinemeyer:2000fa} including the two-loop
contributions to the Higgs boson propagator matrix.
Section 2 summarizes the present theoretical status of Higgs decays into
bottom quark pairs and sets the basis for the resummation, which is
described in Section 3. In Section 4 we analyze the remaining
theoretical uncertainties originating from the SUSY-QCD
corrections in detail for representative MSSM scenarios. In Section 5
we conclude.

\section{Higgs decays into bottom quark pairs}
%        ====================================
\subsection{QCD corrections}
%           ===============
The partial decay widths of the neutral Higgs bosons $\Phi=h,H,A$
into bottom quark pairs, including QCD corrections, can be cast into the
form
\begin{equation}
\Gamma [\Phi \, \to \, b{\overline{b}}] =
\frac{3G_F M_\Phi }{4\sqrt{2}\pi} \overline{m}_b^2(M_\Phi) (g_b^\Phi)^2
\left[ \Delta_{\rm QCD} + \Delta_t^\Phi \right] \, .
\label{eq:gamqcd}
\end{equation}
where regular quark mass effects are neglected. The large logarithmic
part of the QCD corrections has been absorbed in the running
$\overline{\rm MS}$ bottom quark mass $\overline{m}_b(M_\Phi)$ at the
scale of the corresponding Higgs mass $M_\Phi$.  The QCD corrections
$\Delta_{\rm QCD}$ and the top quark induced contributions
$\Delta_t^\Phi$ read as \cite{drees}
\begin{eqnarray}
\Delta_{\rm QCD} & = & 1 + 5.67 \frac{\alpha_s (M_\Phi)}{\pi} + (35.94 -
1.36
N_F) \left( \frac{\alpha_s (M_\Phi)}{\pi} \right)^2 \nonumber \\
& & + (164.14 - 25.77 N_F + 0.259 N_F^2) \left(
\frac{\alpha_s(M_\Phi)}{\pi}
\right)^3 \\
\Delta_t^{h/H} & =& \frac{g_t^{h/H}}{g_b^{h/H}}~\left(\frac{\alpha_s
(M_{h/H})}{\pi}
\right)^2 \left[ 1.57 - \frac{2}{3} \log \frac{M_{h/H}^2}{M_t^2}
+ \frac{1}{9} \log^2 \frac{\overline{m}_b^2
(M_{h/H})}{M_{h/H}^2}\right]\non \\
\Delta_t^A & = & \frac{g_t^A}{g_b^A}~\left(\frac{\alpha_s (M_A)}{\pi}
\right)^2
\left[ 3.83 - \log \frac{M_A^2}{M_t^2} + \frac{1}{6} \log^2
\frac{\overline{m}_b^2 (M_A)}{M_A^2} \right] \non
\end{eqnarray}
where $N_F=5$ active flavours are taken into account. In the
intermediate and large Higgs mass regimes the QCD corrections reduce the
$b\bar b$ decay widths by about 50\% due to the large logarithmic
contributions.

\subsection{SUSY--QCD corrections}
%           =====================
In the MSSM the full SUSY-QCD corrections to the fermionic decay modes
have been computed at NLO \cite{dabel,solaeberl}. In the low-energy limit
$M_\phi,M_Z,m_b \ll m_{\sq_i}, m_{\sgl}$ the results can be cast into
the simple form
\begin{eqnarray}
\Gamma(\phi\to b\bar b) & = & \Gamma_{QCD}(\phi\to b\bar b) \left[1+C_F
C_\phi \frac{\alpha_s}{\pi} \right] \nonumber \\
C_\phi \to C_\phi^{LE} & = & - \kappa_\phi~m_{\sgl}~\mu~\tgb~
             I(m^2_{\sq_1},m^2_{\sq_2},m^2_{\sgl}) \nonumber \\
\kappa_h & = & 1+\frac{1}{\tga~\tgb} \nonumber \\
\kappa_H & = & 1-\frac{\tga}{\tgb} \nonumber \\
\kappa_A & = & 1+\frac{1}{\stgb} \nonumber \\
I(a,b,c) & = & \frac{\displaystyle ab\log\frac{a}{b} + bc\log\frac{b}{c} +
ca\log\frac{c}{a}}{(a-b)(b-c)(c-a)}
\label{eq:sqcdlim}
\end{eqnarray}
$\Gamma_{QCD}(\phi\to b\bar b)$ denotes the QCD-corrected decay width of
Eq.\,(\ref{eq:gamqcd}). It should be noted that NLO-terms involving the
trilinear mixing parameter $A_b$ are absent in Eq.\,(\ref{eq:sqcdlim}).

\section{Effective Lagrangian and resummation}
%        ====================================
\subsection{Construction of the effective Lagrangian}
%           ========================================
The result of Eq.\,(\ref{eq:sqcdlim}) can be derived from the effective
low-energy Lagrangian \cite{CGNW}\footnote{This effective Lagrangian has
been obtained by integrating out the heavy SUSY particles $\sq, \sgl$
and is thus {\it not} restricted to large values of $\tgb$ only. It
should be noted that the scale dependence of the running bottom mass and
Yukawa coupling is purely QCD-initiated, since the heavy SUSY particles
are integrated out at a fixed scale of ${\cal O}(M_{SUSY})$ and thus do
not appear as active partons in the corresponding renormalization group
equations.}
\bea
{\cal L}_{eff} & = & -\lambda_b \overline{b_R} \left[ \phi_1^0
+ \frac{\Delta m_b}{\tgb} \phi_2^{0*} \right] b_L + h.c. \nonumber \\
& = & -m_b \bar b \left[1+i\gamma_5 \frac{G^0}{v}\right] b
-\frac{m_b/v}{1+\Delta m_b} \bar b \left[ g_b^h \left(
1-\frac{\Delta m_b}{\tga~\tgb}\right) h \right. \nonumber \\
& & \hspace*{3cm} \left. + g_b^H \left( 1+\Delta m_b \frac{\tga}{\tgb}\right) H
- g_b^A \left(1-\frac{\Delta m_b}{\stgb} \right) i \gamma_5 A \right] b
\label{eq:leff}
\eea
with
\bea
\Delta m_b & = & \frac{C_F}{2}~\frac{\alpha_s}{\pi}~m_{\sgl}~\mu~\tgb~
I(m^2_{\sq_1},m^2_{\sq_2},m^2_{\sgl}) \nonumber \\
m_b & = & \frac{\lambda_b v_1}{\sqrt{2}} \left[ 1 + \Delta m_b \right]
\nonumber \\
\phi_1^0 & = & \frac{1}{\sqrt{2}}\left[ v_1 + H\coa - h\sia + iA\sib - iG^0\cob
\right] \nonumber \\
\phi_2^0 & = & \frac{1}{\sqrt{2}}\left[ v_2 + H\sia + h\coa + iA\cob + iG^0\sib
\right]
\label{eq:effpar}
\eea
after expansion up to NLO.  The symbol $\phi_1^0 (\phi_2^0)$ denotes the
neutral components of the Higgs doublet coupling to down-(up-)type
quarks. The parameter $\tgb=v_2/v_1$ is defined as the ratio of the two
vacuum expectation values, and $v^2={v_1^2+v_2^2} = 1/{\sqrt{2}G_F}$ is
related to the Fermi constant $G_F$. The would-be Goldstone field $G^0$
is absorbed by the $Z$ boson and generates its longitudinal component.
The SUSY--QCD corrections turn out to be significant for large values of
$\tgb$ and moderate or large $\mu$ values. In order to improve the
perturbative result all terms of ${\cal
O}\left[(\alpha_s\,\mu\,\tgb)^n\right]$ have been resummed \cite{CGNW}.
The correctly resummed effective Lagrangian is given by
Eq.\,(\ref{eq:leff}). The correction \Dmb\ is \textit{non-decoupling} in
the sense that scaling \textit{all} SUSY parameters $m_{\sq_{1,2}},
m_{\sgl}, \mu$ in Eq.~(\ref{eq:effpar}) leaves \Dmb\ invariant. However,
its contribution develops decoupling properties \cite{decoup}, as we
will discuss later on.

Apart from the correction $\Delta m_b$ there is a second class of
potentially large (non-decoupling) contributions
at higher orders which may spoil the 
perturbative reliability of the results: The trilinear mixing parameter
$A_b$ can be of similar size as $\mu\tgb$ as e.g.\,in no-mixing
scenarios of the sbottom particles. In the low-energy limit of
Eq.\,(\ref{eq:sqcdlim}) such terms are absent. However, they arise at
higher orders. In the following we develop an approach to include $A_b$
terms in the resummation of Eq.\,(\ref{eq:leff}). For this purpose we
start from the unrenormalized effective Lagrangian in the low-energy
limit at leading order
\beq
{\cal L}^{LO}_{eff} = -\lambda_b^0 \overline{b_R^0} \phi_1^0 b_L^0 + h.c.
\eeq
Including higher-order corrections in the low-energy limit, the pole mass
$m_b$ of the bottom quark is given by
\beq
m_b = \frac{\lambda_b^0}{\sqrt{2}} v_1 + \Sigma_b(m_b)
\label{eq:bmass}
\eeq
where the self-energy $\Sigma_b(m_b)$ can be decomposed as
\beq
\Sigma_b(m_b) = \frac{\lambda^0_b}{\sqrt{2}}\left[ \Delta_1 v_1 + \Delta_2
v_2 \right] = \frac{\lambda^0_b}{\sqrt{2}} v_1 \left[ \Delta_1 +
\Delta_2\tgb \right]
\eeq
The leading parts in $A_b$ and $\mu$ are finite at NLO,
\bea
\Delta_1 & = & -\frac{C_F}{2}~\frac{\alpha_s}{\pi}~m_{\sgl}~A_b~
I(m^2_{\sq_1},m^2_{\sq_2},m^2_{\sgl}) \nonumber \\
\Delta_2 & = & \frac{C_F}{2}~\frac{\alpha_s}{\pi}~m_{\sgl}~\mu~
I(m^2_{\sq_1},m^2_{\sq_2},m^2_{\sgl}) = \frac{\Delta m_b}{\tgb}
\label{eq:d12}
\eea
Inserting these two expressions in Eq.\,(\ref{eq:bmass}) leads to the
well-known result that the radiative corrections to the bottom mass are
proportional to $A_b-\mu\tgb$, i.e.\,the off-diagonal components of the
sbottom mass matrix of Eq.\,(\ref{eq:sqmassmat}).

\begin{figure}[tb]
\centerline{\includegraphics{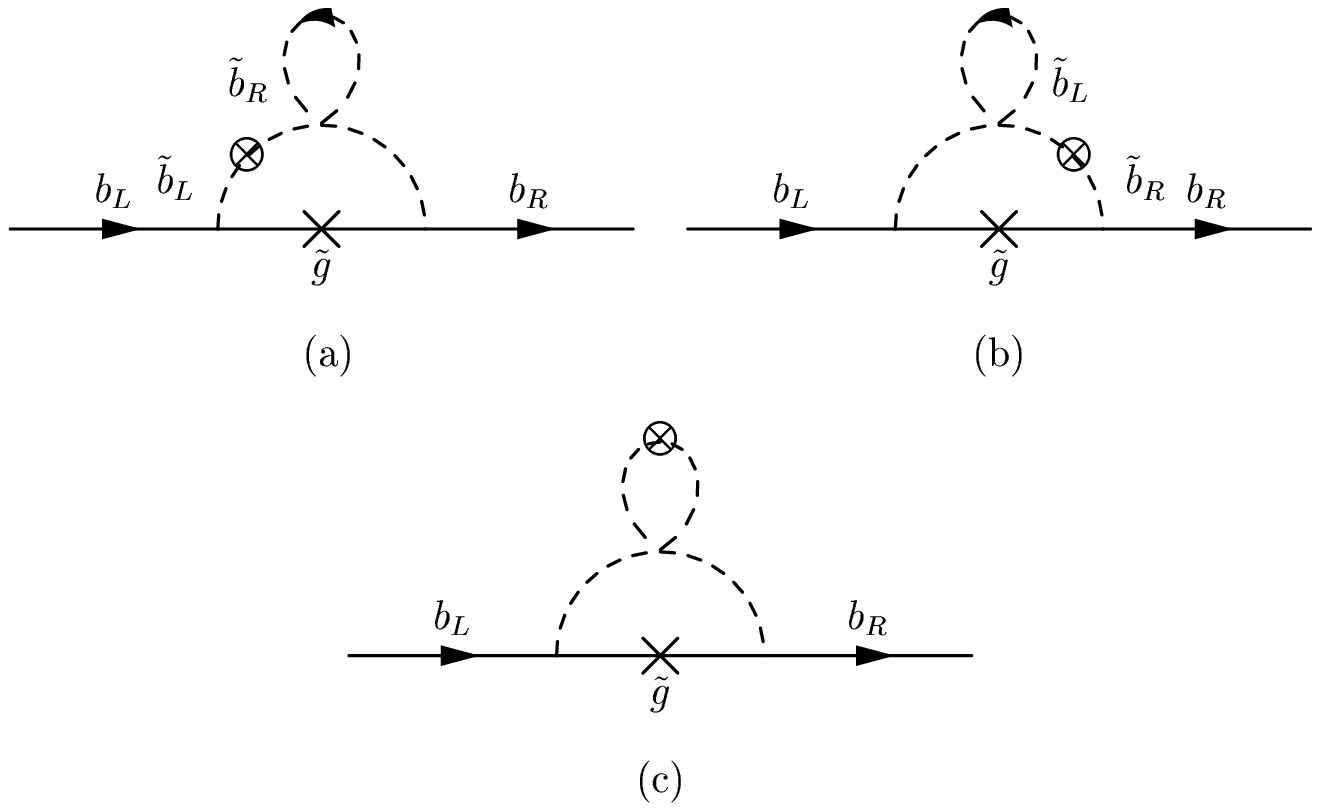}}
\caption{Non-decoupling two-loop contributions to $\Delta_1$ and
$\Delta_2$}
\label{fg:highernondec}
\end{figure}

The structure of the self-energy beyond NLO can be derived from general
arguments based on the asymptotic behaviour of the corresponding
Feynman-diagrams in the low-energy limit. The terms involving $A_b$ or
$\mu$ are generated by mass-insertions in the virtual sbottom
propagators. At NLO the diagrams of Fig.\,\ref{fg:leading}
\figone%
behave asymptotically as\footnote{The functions
$A_0,B_0,C_0,D_0$ denote the usual one-loop scalar integrals for one-,
two-, three- and four-point functions.}
\beq
\alpha_s \lambda_b \left(A_b v_1-\mu v_2\right) m_{\sgl} \times
C_0(0,0;m_{\sq_1}, m_{\sq_2}, m_{\sgl}) \sim \alpha_s m_b m_{\sgl}
\frac{A_b-\mu\tgb}{M_{SUSY}^2}
\label{eq:NLOapp}
\eeq
(for $M_{SUSY}\sim m_{\sq_1}\sim m_{\sq_2}\sim m_{\sgl}$) coinciding
with the explicit results of Eq.\,(\ref{eq:d12}).
At NNLO the leading contributions involving $A_b$ and $\mu$ are
generated by e.g.\,the diagrams of Fig.\,\ref{fg:highernondec}.
The diagrams (a) and (b) behave asymptotically as
\beq
\alpha_s^2 \lambda_b \left(A_b v_1-\mu v_2\right) m_{\sgl} \times
A_0(m_{\sq_i}) \times D_0(0,0,0;m_{\sq_1}, m_{\sq_2}, m_{\sq_j}, m_{\sgl})
\sim \alpha_s^2 m_b m_{\sgl} \frac{A_b-\mu\tgb}{M_{SUSY}^2}
\label{eq:NNLOab}
\eeq
while diagram (c) develops the low-energy behaviour
\beq
\alpha_s^2 \lambda_b \left(A_b v_1-\mu v_2\right) m_{\sgl} \times
B_0(0;m_{\sq_1},m_{\sq_2}) \times C_0(0,0; m_{\sq_i}, m_{\sq_j}, m_{\sgl})
\sim \alpha_s^2 m_b m_{\sgl} \frac{A_b-\mu\tgb}{M_{SUSY}^2}
\label{eq:NNLOc}
\eeq

Thus, the diagrams of Fig.\,\ref{fg:highernondec} contribute to the same
order as the pure QCD corrections to the NLO results and do not generate
leading terms of ${\cal O}(A_b^2)$, ${\cal O}(\mu^2 \stgb)$ nor ${\cal
O}(A_b\mu\tgb)$. This power-counting argument can be applied to all
other two-loop diagrams involving $\mu$ and $A_b$, too. Any further
mass-insertion is suppressed by another power of $m_b/M_{SUSY}$, and is
therefore non-leading.

These arguments can be extended to any perturbative order.
Due to the KLN theorem~\cite{Kinoshita:ur,Lee:is} irreducible
diagrams do not develop power-like divergences in the bottom mass for
$\mb\to0$. Any mass-insertion in the sbottom propagators leads to the
replacement
\bea
\frac{1}{q^2-m_{\sq_i^2}} & \to & \frac{1}{q^2-m_{\sq_1^2}}
\mb(A_b-\mu\tgb) \frac{1}{q^2-m_{\sq_2^2}} \nonumber \\
& \sim & -\frac{m_b(A_b-\mu\tgb)}{M_{SUSY}^2} \frac{1}{q^2-m_{\sq_i^2}}
\nonumber
\eea
Therefore, the low-energy behaviour of the mass-inserted diagram is
modified by an additional power of $m_b(A_b-\mu\tgb)/M_{SUSY}^2$.
Consequently, the diagrams of Fig.\,\ref{fg:highernondec} constitute the
leading contributions in $A_b$ and $\mu\tgb$ at NNLO. These arguments
prove that the results of Eq.\,(\ref{eq:d12}) include all leading powers
of $\alpha_s A_b$ and $\alpha_s \mu\tgb$. This is confirmed by the
explicit two-loop results of Ref.\,\cite{2loop}.

In order to obtain the effective low-energy Lagrangian from the
expression Eq.\,(\ref{eq:bmass}) for the bottom mass, we have to perform
the replacements $v_1\to \sqrt{2}\phi_1^0$ and $v_2\to
\sqrt{2}\phi_2^{0*}$ in the corresponding bottom mass operator. These
replacements lead to the exact interactions with non-propagating Higgs
fields, i.e.\,in the low-energy limit of small Higgs momentum
\cite{let}.  The final expression of the effective Lagrangian can be
cast into the form
\beq
{\cal L}_{eff} = -\lambda_b^0 \overline{b^0_R} \left\{ (1+\Delta_1)
\phi_1^0 + \Delta_2 \phi_2^{0*} \right\} b^0_L + h.c.
\eeq
which differs from previous results by the new factor $(1+\Delta_1)$ in
front of $\phi_1^0$.
This expression has to be matched with the renormalized low-energy Lagrangian
\beq
{\cal L}_{eff} = - \lambda_b \overline{b_R} \left\{ \phi_1^0 +
\frac{\Delta_b}{\tgb} \phi_2^{0*} \right\} b_L + h.c.
\label{eq:Lefffinal}
\eeq
yielding the relations\footnote{It should be noted that the bottom
wave-function renormalization constants do not contain any leading
non-decoupling contribution in $A_b$ and $\mu$. Moreover, it should be
emphasized that the combination $A_b-\mu\tgb$ only appears in the
definition of the bottom mass, while $A_b$ and $\mu\tgb$ contribute in a
different way to the bottom Yukawa coupling and Higgs decay processes.}
\bea
\lambda_b & = & \lambda_b^0 (1+\Delta_1) \nonumber \\
\Delta_b & = & \frac{\Delta_2 \tgb}{1+\Delta_1}
= \frac{\Delta m_b}{1+\Delta_1}
\label{eq:delb}
\eea
Thus all terms of ${\cal O}[(\alpha_s/M_{SUSY})^n (\mu\tgb)^m
A_b^{n-m}]$ are resummed by means of the simple replacement
\beq
\Delta m_b \to \frac{\Delta m_b}{1+\Delta_1}
\label{eq:dmbdel1}
\eeq
in the effective Lagrangian of Eq.\,(\ref{eq:leff}). This proof confirms
and extends the resummation presented in Ref.\,\cite{CGNW} and explains
the absence of any $A_b$ terms in Eq.\,(\ref{eq:sqcdlim}) in terms of
a clear physical interpretation: the leading $A_b$ terms are absorbed in
the definition of the effective Yukawa coupling $\lambda_b$ in the
low-energy effective Lagrangian. In a Feynman diagrammatic approach this
corresponds to a cancellation of the $A_b$ terms in the bottom-mass
counterterms and the genuine irreducible three-point diagrams. This
cancellation is exact at zero-momentum transfer, but a mild dependence
on $A_b$ appears when keeping all external momenta on-shell due to the
momentum dependence of the one-particle-irreducible (1PI) three-point
functions.

The final results for the resummed partial decay widths can be
cast into the form [see
Eqs.\,(\ref{eq:gamqcd}--\ref{eq:sqcdlim})]\footnote{In order to
avoid an artificial singularity in $\Gamma(h\to b\bar b)$ for
vanishing $\alpha$ the remainder proportional to $(C_\phi -
C_\phi^{LE})$ is multiplied by the unresummed Yukawa coupling
$g_b^\phi$.}
\begin{equation}
\Gamma [\Phi \, \to \, b{\overline{b}}] =
\frac{3G_F M_\Phi }{4\sqrt{2}\pi} \overline{m}_b^2(M_\Phi)
\left[ \Delta_{\rm QCD} + \Delta_t^\Phi \right] \tilde g_b^\Phi \left[
\tilde g_b^\Phi + g_b^\Phi (C_\phi - C_\phi^{LE}) \frac{\alpha_s}{\pi}
\right]
\label{eq:gamfin}
\end{equation}
with the resummed couplings [see
Eqs.\,(\ref{eq:leff},\ref{eq:Lefffinal},\ref{eq:delb})]
\bea
\tilde g_b^h & = & \frac{g_b^h}{1+\Delta_b} \left(
1-\frac{\Delta_b}{\tga~\tgb}\right) \nonumber \\
\tilde g_b^H & = & \frac{g_b^H}{1+\Delta_b} \left(
1+\Delta_b \frac{\tga}{\tgb}\right) \nonumber \\
\tilde g_b^A & = & \frac{g_b^A}{1+\Delta_b} \left(
1-\frac{\Delta_b}{\stgb}\right)
\eea

\subsection{Validity of the low-energy approximation}
%           ========================================
The expression in Eq.~(\ref{eq:Lefffinal}) resums the terms of ${\cal
O}[(\alpha_s/M_{SUSY})^n (\mu\tgb)^m A_b^{n-m}]$ to all orders in
perturbation theory. However, there are other kinds of non-decoupling
terms in the 1PI self-energies, as can be inferred already from the NNLO
expressions of Eqs.~(\ref{eq:NNLOab},\ref{eq:NNLOc}). The question about
the numerical size of these non-leading terms arises, and whether the
NNLO resummation is necessary in practical applications.
Eqs.~(\ref{eq:NLOapp}--\ref{eq:NNLOc}) imply that the irreducible NNLO
corrections $\Delta_1^{(2)}$ and $\Delta_2^{(2)}$ to the selfenergy are
of the order of $\Delta_{\{1,2\}}^{(2)} \sim \alpha_s \Delta_{\{1,2\}}$,
while the reducible diagrams contribute as $(\Delta_{\{1,2\}})^2$.  For
the irreducible diagrams to be dominant compared to the reducible ones,
the condition $(\Delta_{\{1,2\}})^2 \lsim |\Delta_{\{1,2\}}^{(2)}| \sim
\alpha_s |\Delta_{\{1,2\}}|$ has to be fulfilled, i.e.\,
$|\Delta_{\{1,2\}}| \lsim \alpha_s\sim {\cal O}(10\%)$. Therefore, the
scenarios with the NNLO 1PI being dominant lead to
$|\Delta_{\{1,2\}}^{(2)}| \lsim {\cal O}(1\%)$, so that the NLO
corrections are small, and the size of the NNLO corrections is of the
same order as the deviation of the full results from the zero-momentum
approximation.  This argument can be extended to higher orders in
perturbation theory. At the $n$-loop level the non-decoupling 1PI
diagrams originate from a single vacuum insertion (analogous to the
diagrams of Fig.~\ref{fg:highernondec}) which are of ${\cal
O}(\alpha_s^n \mb \mg (A_b -\mu\tgb)/M_{SUSY}^{2})\simeq \alpha_s^{n-1}
\Delta_{\{1,2\}}$. Hence, they are negligible, because either they are
much smaller than the $n$-loop reducible contribution, or the numerical
value of the leading corrections is small already at NLO.

The trilinear mixing parameter $A_b$ cannot be much larger than
$M_{SUSY}$, since otherwise the color and charge symmetries would be
broken \cite{ccb}. Thus, the contribution $\Delta_1$ of
Eq.\,(\ref{eq:d12}) reaches maximal values of ${\cal O}(10\%)$, while the
term $\Delta m_b$ can be larger by an order of magnitude.

\begin{figure}[hbtp]
 \setlength{\unitlength}{1cm}
 \centering
 \begin{picture}(15,20.0)
  \put(2.0, 6.0){\epsfxsize=12cm \epsfbox{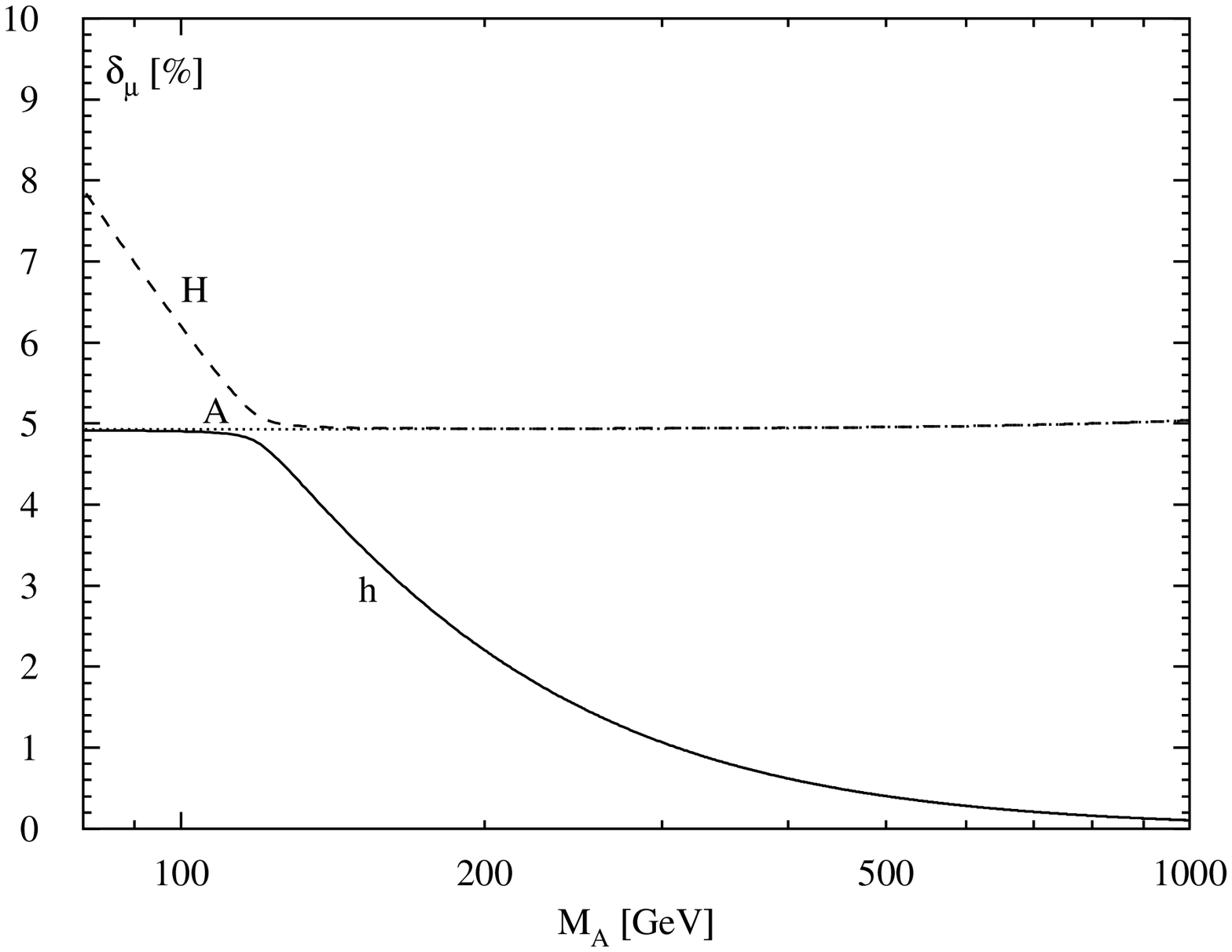}}
  \put(2.0,-4.3){\epsfxsize=12cm \epsfbox{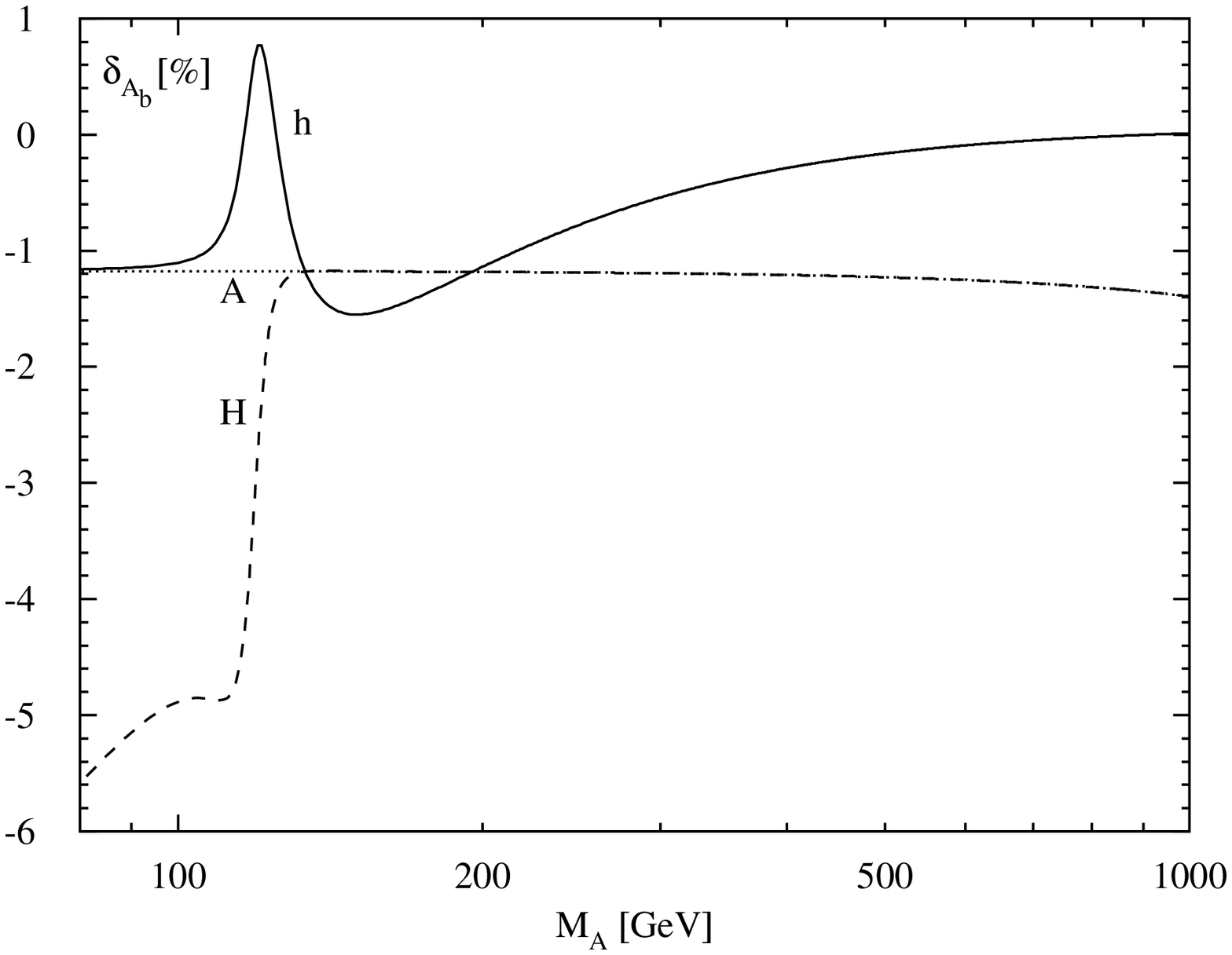}}
  \put(13.0,19.0){\bf (a)}
  \put(13.0, 8.7){\bf (b)}
 \end{picture}
\caption[]{\it \label{fg:diff} Relative corrections due to (a) the
SUSY--QCD corrections including the resummation of $\Delta m_b$ of
Eq.~(\ref{eq:effpar}) and (b) due to $\Delta_1$ of Eq.~(\ref{eq:d12}) as
a function of the pseudoscalar mass $M_A$ for all neutral Higgs bosons.
The relative corrections are normalized to the QCD-corrected decay
widths $\Gamma_{QCD}(\phi\to b\bar b)$ of Eq.\,(\ref{eq:gamqcd}) in both
cases.}
\end{figure}

In Fig.\,\ref{fg:diff} we compare the relative NLO corrections including
the resummation of $\Delta m_b$ with the novel NNLO contributions
$\Delta_1$ of Eq.\,(\ref{eq:effpar}) as a function of the pseudoscalar
Higgs mass $M_A$ for all three neutral Higgs states in the following
MSSM scenario with large $A_b$:
\bea
\tgb & = & 30 \nonumber \\
M_{\tilde Q} & = & 2~{\rm TeV}\nonumber \\
M_{\sgl} & = & 1.6~{\rm TeV} \nonumber \\
A_t & = & \mu \cot \beta \nonumber \\
A_b & = & -\mu\,\tgb \nonumber \\
\mu & = & -150~{\rm GeV}
\eea
The relative corrections are normalized to the QCD-corrected decay
widths $\Gamma_{QCD}(\phi\to b\bar b)$ of Eq.\,(\ref{eq:gamqcd}) in both
cases. While the $\Delta m_b$ effects are of ${\cal O}(10\%)$ and thus
of moderate size, the novel $\Delta_1$ contributions turn out to be of
${\cal O}(1\%)$ apart from the small heavy scalar Higgs mass range,
where they can reach a similar magnitude as the $\Delta m_b$ terms. This
particular scenario, however, has to be considered as an extreme case.
In general the $\Delta_1$ terms are small, confirming the previous
qualitative discussion.

\section{Numerical Results}
%        =================
The numerical analysis of the neutral Higgs boson decays into bottom
quark pairs is performed for the 'small $\alpha_{eff}$' MSSM scenario
\cite{bench} as a representative case:
\bea
\tgb & = & 30 \nonumber \\
M_{\tilde Q} & = & 800~{\rm GeV}\nonumber \\
M_{\sgl} & = & 500~{\rm GeV} \nonumber \\
M_2 & = & 500~{\rm GeV} \nonumber \\
A_b = A_t & = & -1.133~{\rm TeV} \nonumber \\
\mu & = & 2~{\rm TeV}
\eea
We use the RG-improved two-loop expressions of Ref.\,\cite{rgi}.  The
bottom quark pole mass has been chosen to be $M_b=4.62$ GeV, which
corresponds to a $\MS$ mass $\overline{m}_b(\overline{m}_b)=4.28$~GeV.
The strong coupling constant has been normalized to
$\alpha_s(M_Z)=0.119$.

\begin{figure}[hbt]
 \setlength{\unitlength}{1cm}
 \centering
 \begin{picture}(15,11.0)
  \put(1.0,-4.5){\epsfxsize=13cm \epsfbox{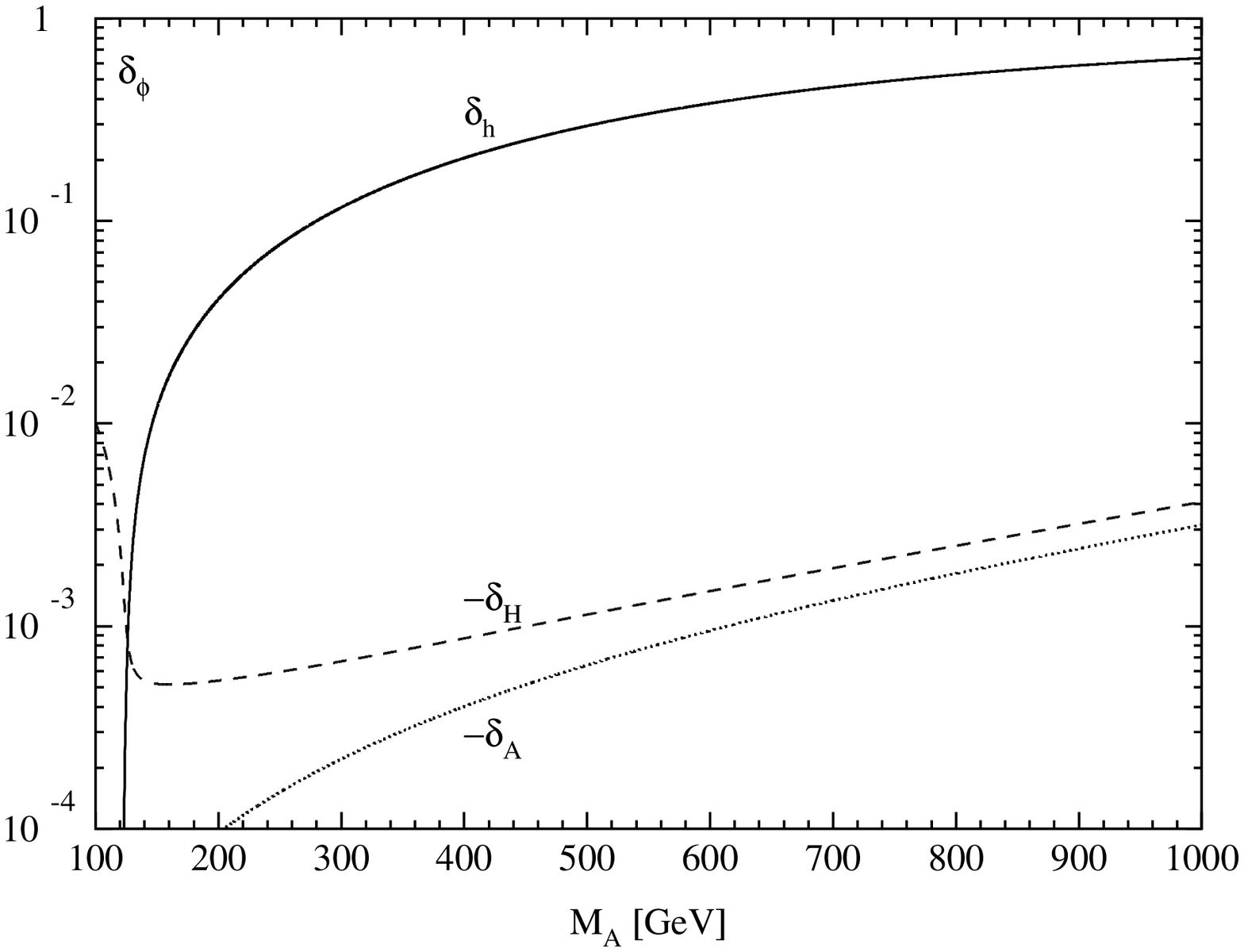}}
 \end{picture}
\caption[]{\it \label{fg:approx} Relative deviations $\delta_\phi$ of the
approximate low-energy one-loop result from the full NLO expression as a
function of the pseudoscalar mass $M_A$ in the 'small $\alpha_{eff}$'
scenario (a) for all neutral Higgs bosons. For the heavy scalar and
pseudoscalar Higgs bosons the deviations are negative. The values shown
have to be changed in sign.}
\end{figure}

The resummation effects discussed in the previous section have been
derived in the low-energy limit $M_\phi^2,M_Z^2,m_b^2\ll M_{SUSY}^2$.
The question arises, how reliable this approximation works in
phenomenological applications. In particular, the magnitude of ${\cal
O}(M_\phi^2/M_{SUSY}^2, M_Z^2/M_{SUSY}^2, m_b^2/M_{SUSY}^2)$ terms
matters for sizeable masses of the low-energy particles. This can be
tested explicitly by comparing the approximate results of
Eq.\,(\ref{eq:sqcdlim}) with the full one-loop result. A typical example
is depicted in Fig.\,\ref{fg:approx} for the 'small $\alpha_{eff}$'
scenario, where the relative difference between the full and approximate
one-loop contributions [see Eq.\,(\ref{eq:sqcdlim})]
\beq
\delta_\phi = \frac{C_\phi - C_\phi^{LE}}{C_\phi}
\eeq
is presented for all neutral Higgs particles as a function of the
pseudoscalar Higgs mass $M_A$. It is clearly visible that the
approximation turns out to be sufficient for the heavy neutral Higgs
particles $H,A$, but fails for the light scalar Higgs boson $h$ in the
decoupling limit \cite{petra}. 
However, in the decoupling limit the size of the
approximate SUSY--QCD corrections strongly decreases, since $\tga \to
-1/\tgb$ and thus
\beq
\frac{1}{1+\Delta m_b} \left( 1-\frac{\Delta m_b}{\tga~\tgb} \right) \to
1
\eeq
so that the SUSY--QCD corrections become negligible. Due to this
behaviour the low-energy approximation is sufficient for most
phenomenological applications. This also explains the failure of the
approximation in this case: the large non-decoupling contributions from
\Dmb\ cancel to a large extent in the lightest Higgs boson couplings,
leaving a small remainder of the same order as the non-leading
contributions. On the other hand, this cancellation does not occur for
the heavy Higgs bosons, and the effective Lagrangian approach yields a
good approximation.

There are two basic sources of systematic uncertainties
originating from the SUSY-QCD contributions: {\it (i)} The
MSSM masses and couplings involved in the NLO SUSY--QCD corrections will
only be known with a sizeable uncertainty at the LHC, while future
$e^+e^-$ linear colliders in the 500 GeV to 1 TeV range will enable
precision measurements of the SUSY masses and couplings. These errors in
the input parameters generate systematic uncertainties for the
prediction of the partial decay widths. {\it (ii)} Due to missing higher
order results the scale dependence of the strong coupling constant
$\alpha_s$ will not be compensated. The scale variation yields an
estimate of the purely theoretical SUSY-QCD uncertainty, which
will be analyzed quantitatively in this section.\footnote{The
electroweak contributions introduce additional uncertainties, which are
not taken into account. They provide contributions to $\Delta_1$ and
$\Delta_2$ in addition to the SUSY-QCD part.}

The central scale $\mu_0$ of the strong coupling constant appearing in
the SUSY--QCD corrections will be chosen as the average mass of the
involved SUSY particles, i.e.
\beq
\mu_0 = \frac{m_{\sq_1}+m_{\sq_2}+m_{\sgl}}{3}
\eeq
In order to estimate the residual scale dependence the scale of
$\alpha_s$ will be varied between $\mu_0/3$ and $3\mu_0$.  The usual QCD
corrections have been included up to the three-loop order so that the
residual purely QCD-induced scale dependence ranges below the per-mille
level and can thus safely be neglected.

\begin{figure}[hbtp]
 \setlength{\unitlength}{1cm}
 \centering
 \begin{picture}(15,21.0)
  \put(3.0,11.4){\epsfxsize=8.2cm \epsfbox{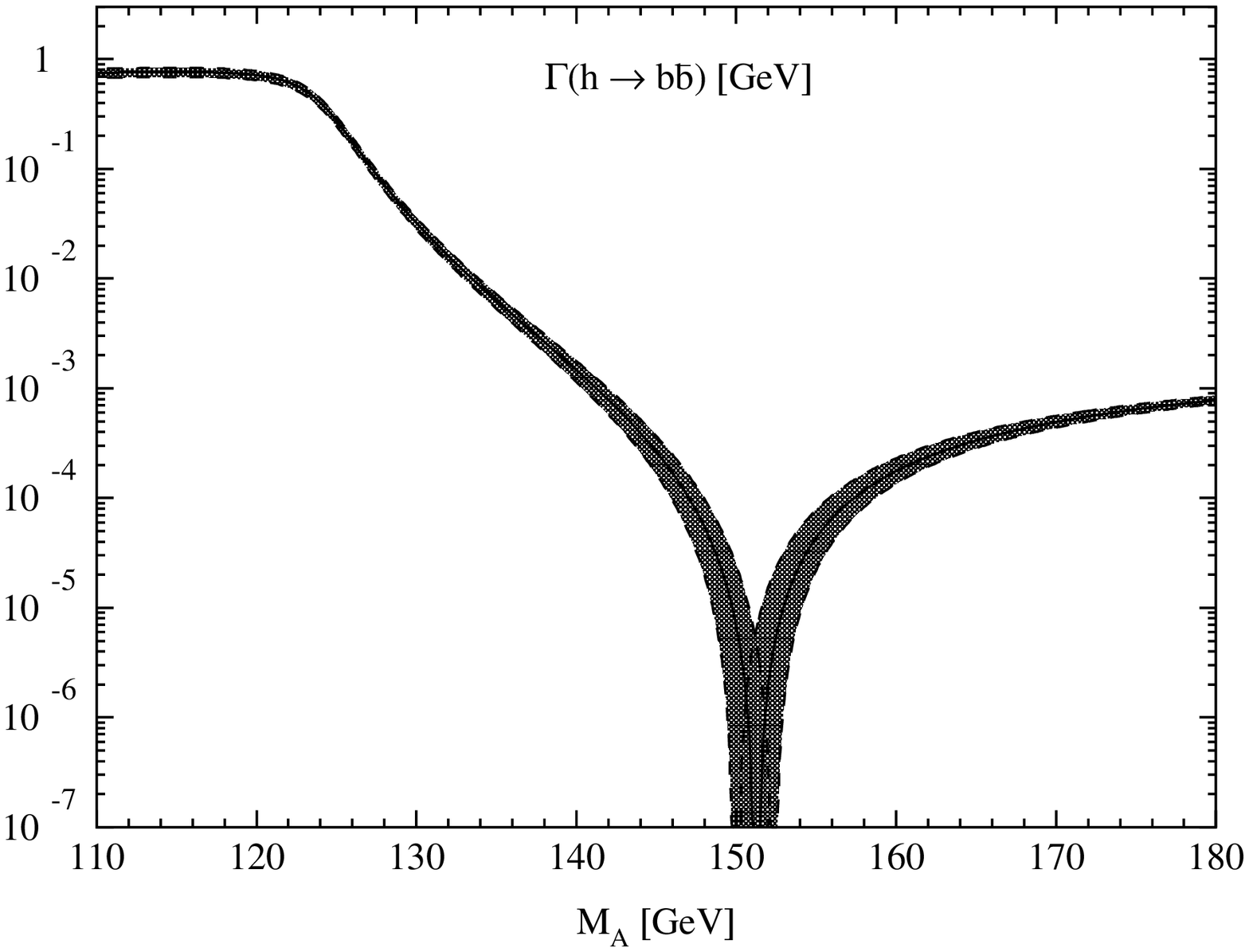}}
  \put(3.0, 4.2){\epsfxsize=8.2cm \epsfbox{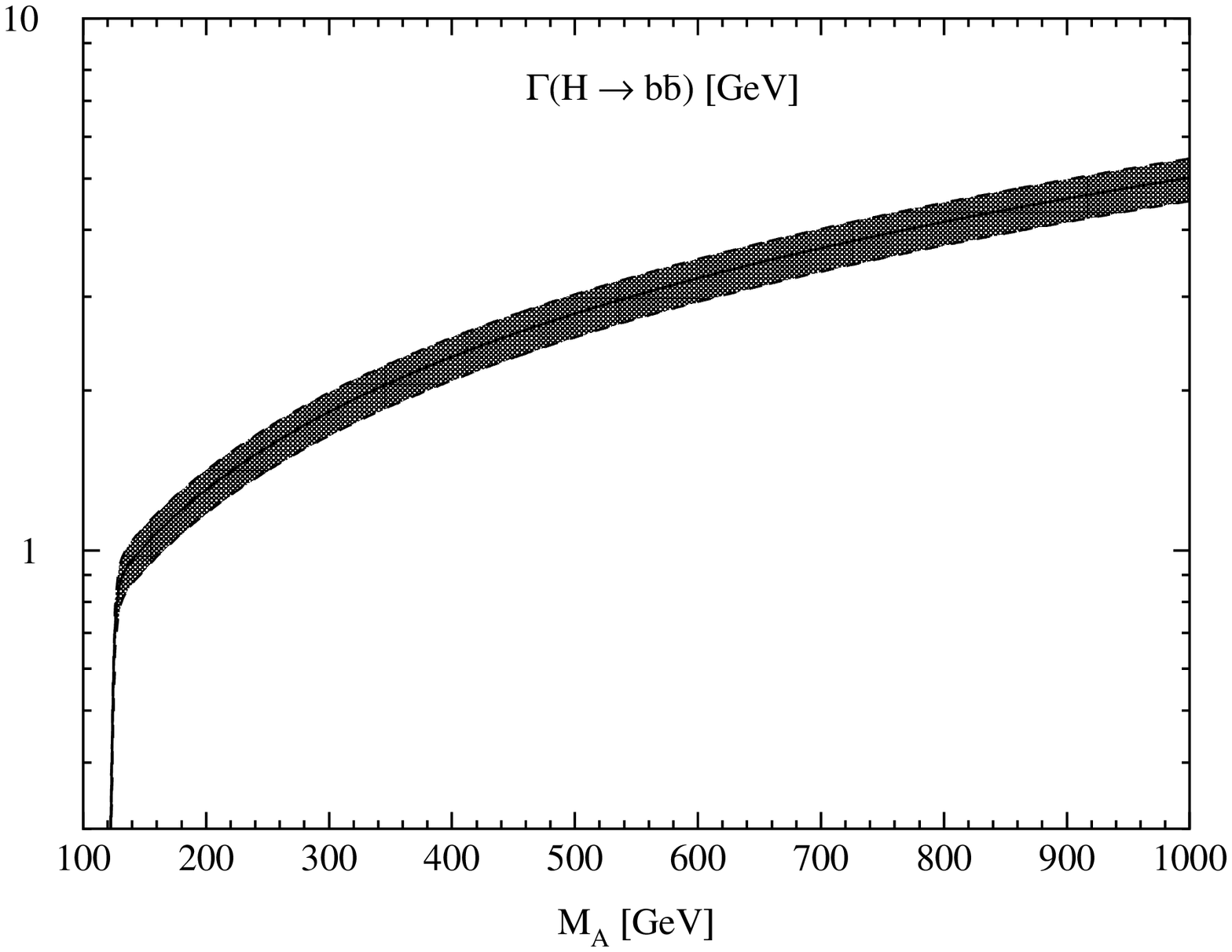}}
  \put(3.0,-3.0){\epsfxsize=8.2cm \epsfbox{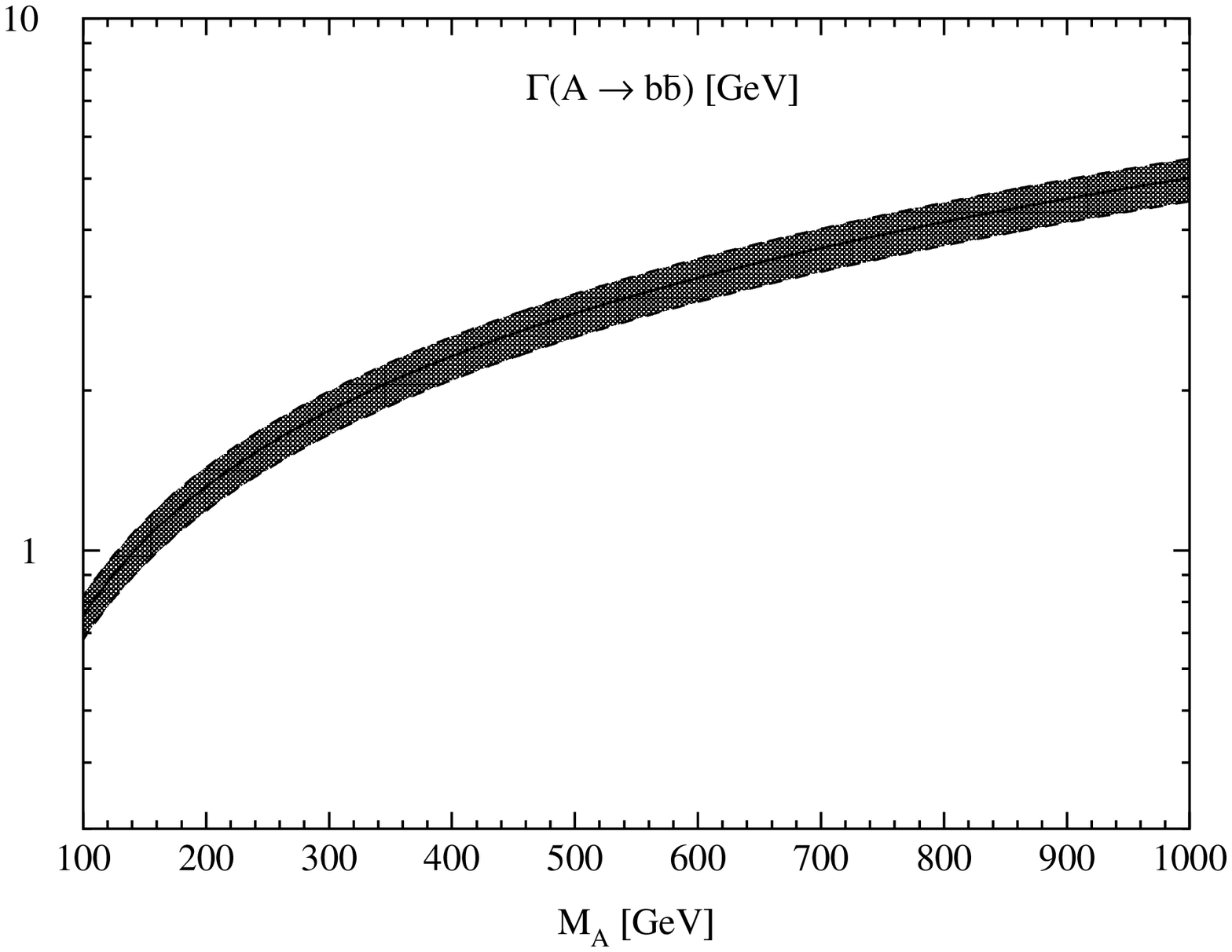}}
  \put(5.5,20.3){\bf (a)}
  \put(5.5,13.1){\bf (b)}
  \put(5.5, 5.9){\bf (c)}
 \end{picture}
\caption[]{\it \label{fg:parwid} Partial decay widths $\Gamma(\phi\to
b\bar b)$ of (a) the light scalar, (b) the heavy scalar and (c) the
pseudoscalar Higgs boson in the 'small $\alpha_{eff}$' scenario. The
shaded bands reflect the uncertainties due to the scale choice of the
strong coupling constant $\alpha_s$.}
\end{figure}
The results for the partial decay widths are shown in
Fig.\,\ref{fg:parwid}a for the light scalar Higgs boson, in
Fig.\,\ref{fg:parwid}b for the heavy scalar Higgs boson and in
Fig.\,\ref{fg:parwid}c for the pseudoscalar Higgs boson.  These results
include the QCD corrections up to NNNLO of Eq.\,(\ref{eq:gamqcd}) and
the full NLO SUSY--QCD corrections of Eq.\,(\ref{eq:sqcdlim}) with the
resummation of the leading $\Delta m_b$ and $\Delta_1$ terms according
to Eqs.\,(\ref{eq:leff},\ref{eq:dmbdel1}).  It can clearly be inferred
from these figures that the remaining uncertainties due to the scale
choice are typically of the order of 10\%. However, they are
significantly enhanced in regions where the SUSY--QCD corrections become
large, as in the 'small $\alpha_{eff}$'--scenario, which develops a
strongly suppressed partial decay width $\Gamma(h\to b\bar b)$ for
pseudoscalar masses $M_A\sim 150$ GeV\footnote{The explicit value of the
pseudoscalar mass where the Yukawa coupling vanishes depends strongly on
the included higher-order corrections.}.  This, however, corresponds
only to a tiny region in the light scalar Higgs mass $M_h$ close to its
upper limit for large $M_A$ within the 'small
$\alpha_{eff}$'--scenario. The theoretical uncertainties turn out to be
large at $M_A\sim 150$ GeV.

\begin{figure}[hbtp]
 \setlength{\unitlength}{1cm}
 \centering
 \begin{picture}(15,21.0)
  \put(3.0,11.4){\epsfxsize=8.2cm \epsfbox{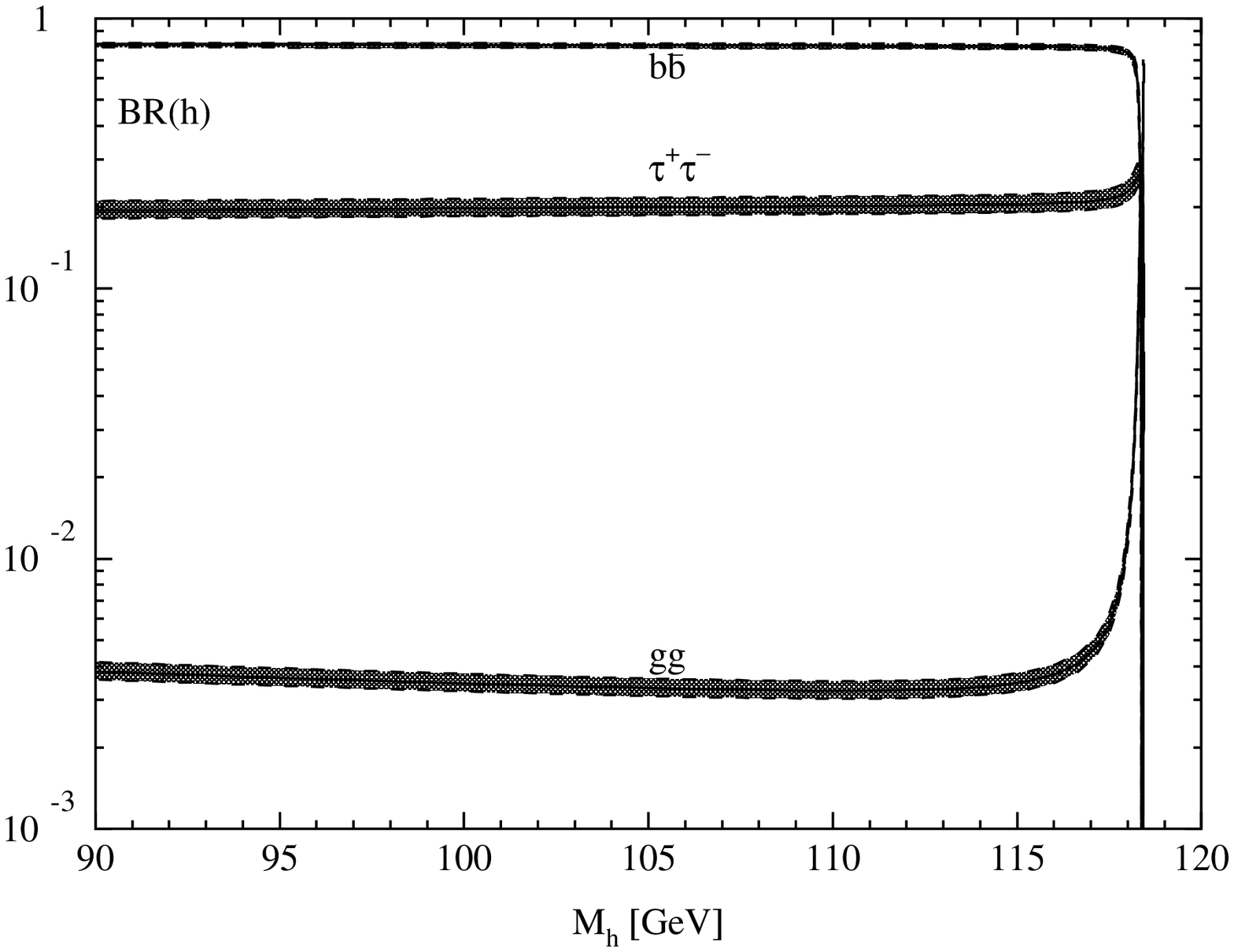}}
  \put(3.0, 4.2){\epsfxsize=8.2cm \epsfbox{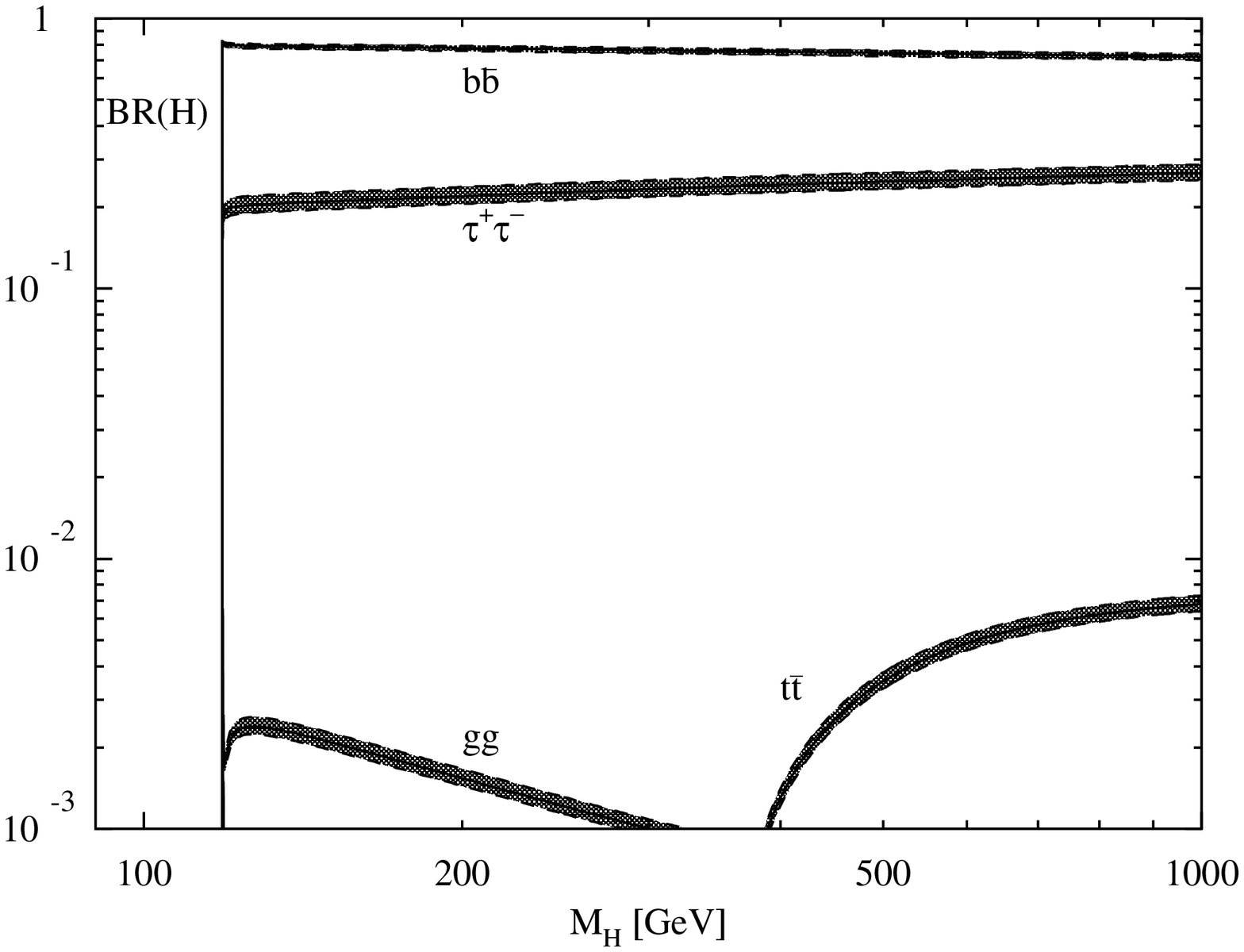}}
  \put(3.0,-3.0){\epsfxsize=8.2cm \epsfbox{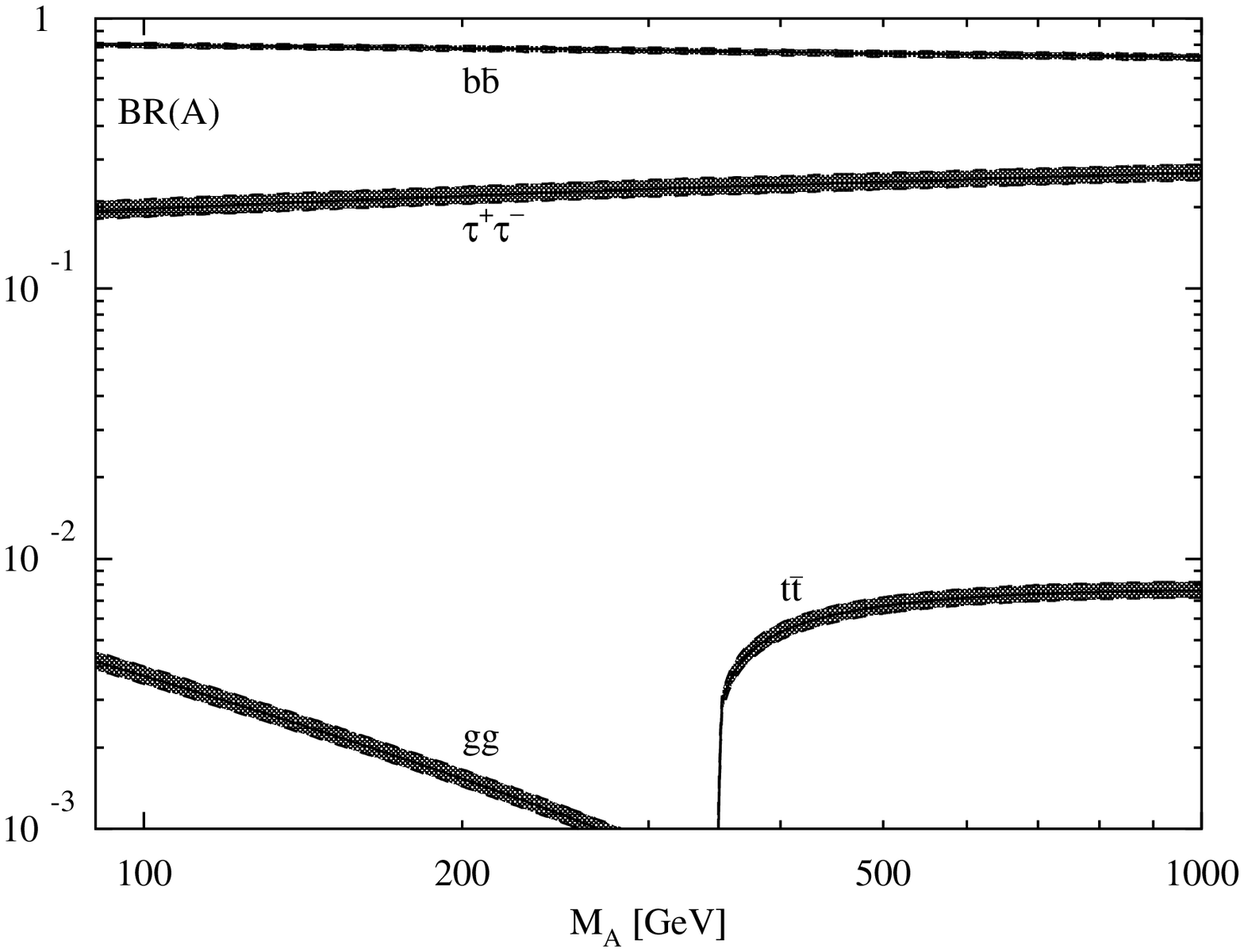}}
  \put(3.5,17.8){\bf (a)}
  \put(3.5,10.6){\bf (b)}
  \put(3.5, 3.4){\bf (c)}
 \end{picture}
\caption[]{\it \label{fg:br} Branching ratios of (a) the light
scalar, (b) the heavy scalar and (c) the pseudoscalar Higgs boson in the
'small $\alpha_{eff}$' scenario. The shaded bands reflect the
uncertainties due to the scale choice of the strong coupling constant
$\alpha_s$.}
\end{figure}

The uncertainties in the partial decay widths $\Gamma(\phi\to b\bar b)$
translate into systematic errors in the corresponding branching ratios.
They are depicted in Figs.\,\ref{fg:br}a--c for the three neutral Higgs
bosons. These results have been obtained with the program HDECAY
\cite{hdecay} after including the results obtained in this analysis.
Since the partial decay into $b\bar b$ pairs is dominant in nearly the
entire Higgs mass ranges, its uncertainty due to the scale choice above
reduces to a level of ${\cal O}(1\%)$. However, the scale dependence of
$\Gamma(\phi\to b\bar b)$ develops significant systematic errors in the
non-leading branching ratios into $\tau^+\tau^-$, gluon and $t\bar t$
pairs. These can reach a level of ${\cal O}(10\%)$ and are larger than
the expected experimental accuracy at future $e^+e^-$ linear colliders,
which clearly calls for a NNLO calculation of the SUSY--QCD part. These
theoretical errors have to be added to the uncertainties due to
inaccuracies of the input parameters as presented in \cite{hbrqcd}
and the theoretical errors of the Higgs masses and couplings
\cite{Heinemeyer:1999zf}\footnote{The uncertainties due to the Higgs
masses will be eliminated to a large extent, once they
will be measured directly in future experiments.}. They constitute a
significant source of uncertainty. An analogous analysis is required for
the theoretical uncertainties due to the SUSY-electroweak corrections
beyond NLO. However, this is beyond the scope of our paper.

\section{Conclusions}
%        ===========

In this paper we have reanalyzed the neutral scalar Higgs decays into
$b\bar b$ pairs in the MSSM with particular emphasis on the SUSY--QCD
corrections and their theoretical uncertainties. We have extended the
resummation of large non-decoupling SUSY--QCD corrections of ${\cal
O}(\alpha_s \mu \tgb/M_{SUSY})$ by the inclusion of non-decoupling terms
of ${\cal O}(\alpha_s A_b/M_{SUSY})$ which have not been taken into
account in previous analyses. We have shown that these terms are absent
at NLO in the effective Lagrangian but arise at NNLO and beyond. This
can easily be traced back to the renormalization of the bottom Yukawa
coupling in the low-energy limit, where the heavy SUSY particles are
integrated out. We have obtained the important result that these novel
contributions hardly affect the theoretical predictions for the partial
decay widths into $b\bar b$ pairs so that they do not endanger the
reliability of the perturbative result in contrast to the leading terms
of ${\cal O}(\alpha_s \mu \tgb/M_{SUSY})$.

We investigated the remaining theoretical uncertainties generated by the
SUSY--QCD corrections quantitatively. While the theoretical errors of
the partial decay widths $\Gamma(\phi\to b\bar b)$ turn out to be of
${\cal O}(10\%)$, this effect cancels to a large extent in the branching
ratios $BR(\phi\to b\bar b)$ due to its dominance. It appears, however,
as a sizeable increase in the systematic uncertainties of the
non-leading branching ratios into $\tau^+\tau^-$, gluon and $t\bar t$
pairs, which appear to be larger than the anticipated experimental
accuracies at future linear $e^+e^-$ colliders. This clearly calls for a
NNLO calculation of the SUSY--QCD part, which is beyond the scope of
this work. \\

\noindent
{\bf Acknowledgements.} \\
J.G. would like to thank the Max-Plank-Institute f\"ur Physik in Munich
for their kind hospitality. We are indebted to G. Weiglein for useful
discussions. We are grateful to P.\,Zerwas for carefully reading the
manuscript.

%\newpage

\end{document}